\documentclass[a4paper]{article}
\usepackage{graphicx,amsmath}
\usepackage[T1]{fontenc}
\usepackage[latin2]{inputenc}

\usepackage{latexsym,amsfonts,amssymb,amsmath,amsthm}
\usepackage{fancyhdr}
\usepackage{lmodern}
\DeclareMathAlphabet{\mathbk}{OT1}{cmr}{bx}{sl}

\usepackage{eufrak}

\newtheorem{all}{Proposition}[section]

\newcommand\nc{\newcommand}
\nc\ee{\mathbk e}
\nc\e{\EuFrak e}
\nc\E{\mathbk E}
\nc\aS{\mathtt S}
\nc\Sd{\mathbf S}
\nc\la{\lambda}
\nc\Ll{\mathbk L} \nc\Pt{\mathbk P}\nc\Mt{\mathbk M}
\newcommand\1{\boldsymbol 1}\nc\0{\boldsymbol 0}
 \newcommand\rr{\mathbb R}
\newcommand\vf{\varphi} \newcommand\T{\mathbk T}

\newcommand\dd{\mathbk d}

\nc\beq{\begin{equation}}\nc\enq{\end{equation}}
\nc\x{\mathbk x} \nc\y{\mathbk y}\nc\z{\mathbk z}
\nc\ts{\mathbk t}
\nc\intl{\int\limits}
\nc\tm{\mathrm{tm}}
\nc\n{\mathbk n}\nc\nb{\mathbk n}
\nc\aI{\mathrm I}
\nc\pt{\partial}
\nc\rs{\mathbk r}
\nc\vv{\mathbk v}\nc\hh{\mathbk h}
\newcommand\aM{\mathrm M}\newcommand\M{\mathbf M}
\nc\I{\mathbb T}
\nc\uu{\mathbk u}\nc\U{\mathbk U}
\nc\B{\mathbk B} \nc\F{\mathbk F}
\nc\ii{\mathbk i}
\nc\ran{\mathrm{Ran}}
\nc\jj{\EuFrak j}\nc\ji{\mathbk j}
\nc\Dd{\mathrm D}
\nc\s{\mathbk s}
\nc\q{\mathbk q}
\nc\ad{\mathbk a}
\nc\Supp{\mathrm{Supp}}
\nc\R{\mathbk R}
\nc\Ord{\mathrm{Ordo}}\nc\ord{\Ord(\rho)}
\nc\app{\approx}
\nc\otm{\otimes}
\nc\sk{\smallskip}
\nc\dZ{\mathbk Z}
\nc\A{\mathbk A}
\nc\Ef{\boldsymbol{\mathcal E}}
\nc\Bf{\boldsymbol{\mathcal B}}
\nc\Lf{\boldsymbol{\mathcal L}}
\nc\Ff{\boldsymbol{\mathcal F}}
\nc\Tf{\boldsymbol{\mathcal T}}
\nc\Kf{\mathcal K}
\nc\f{\mathbk f}
\nc\Dp{\mathcal D}
\nc\Ht{\mathbk H}\nc\Dt{\mathbk D}
\nc\mqed{\blacksquare}

\begin{document}

\centerline{\Large{\bf On the Radiation Reaction Force}}
\medskip

\centerline{T. Matolcsi\footnote{Department of Applied Analysis and Computational Mathematics,
E\"otv\"os Lor\'and University, Budapest, Hungary}}

\begin{abstract}

The usual radiation self-force of a point charge is obtained in a mathematically exact way and it is pointed 
out to that this does not call forth that the spacetime motion of a point charge obeys 
the Lorentz--Abraham--Dirac equation.\end{abstract}
\bigskip

\part*{Introduction}
\medskip

The equation of spacetime motion of a point charge  under the action of an external force is an old problem of electromagnetism.
The original work of Abraham (treated e.g. in \cite{Jackson}, Ch. 16.3) and then Dirac's formulation 
(treated e.g. in \cite{Groot}, Ch III.3) suffers from a hardly acceptable mass renormalization according to which the finite 
mass of a particle is the difference of the positive infinite electric mass and the negative infinite mechanical mass. Other, 
quick deductions based on Larmor's formula and some other `natural' requirements (see \cite{Jackson}, 
Ch 16.2) are not convincing either. 
Moreover, the LAD equation  admits physically untenable run-away solutions. 
There are some attmepts to exclude such non-physical solutions by imposing extra conditions.
(e.g. \cite{Spohnc}, \cite{lordir}).
There are a number of articles (e.g.\cite{Bay-Hus}, \cite{Rohrlich1}, \cite{Rohrlich2},
\cite{Gral-Har-Wald}, \cite{Spohn}) considering various continuum charge distributions 
instead of a point charge and then taking or not the limit to a point.
Though casting more and more new light on the problem and its possible solution, neither way seems 
satsifactory.

A recent article \cite{Bild-Deck-Ruhl} -- though its result rests on an 
erroneous basis, see \cite{matvan} -- points out clearly 
to Dirac's unjustified applications of Gauss--Stokes theorem, Taylor expansion and the limit procedure when an 
extended charge is shrunk to a point.

In some papers (e.g. \cite{Taylor}, \cite{Rowe},\cite{Gral-Har-Wald}) special applications of 
Distribution Theory can be found when treating point charges.

{\it In the present paper, by a systematic use of Distribution Theory, it is shown that not a mathematically 
incorrect derivation of the radiation reaction force  but a 
physical misapprehension is the source of why the LAD equation does not work well}.

The coordinate-free formulation of spacetime expounded in \cite{Matolcsi} is used which makes 
formulae shorter and more easily comprehensible; find enclosed  
a brief summary of the fundamental notions and some special notations.

Spacetime points and spacetime vectors (which are often confused in coordi\-nates) are distinguished:

-- \ $\aM$, mathematically an affine space, is the set of spacetime points, its elements are denoted by
normal letters, $x$, $y$ etc.,

-- \ $\M$,  mathematically a vector space, is the set of spacetime vectors, its elements are denoted by
boldface letters, $\x$,  $\y$ etc.,

-- \ the set of time periods is $\I$,

-- \ the exact treatment requires the tensorial quotients of $\M$ by $\I$, here we do not refer to them explicitely, 

-- \ the Lorentz product of the vectors $\x$ and $\y$ is denoted by $\x\cdot\y$ (which is $x_ky^k$ in coordinates);
$\x$ is timelike if $\x\cdot\x<0$ (which means in coordinates the signature $(-1,1,1,1)$ of the Lorentz form),

-- \ an absolute velocity is a futurelike vector $\uu$ for which $\uu\cdot\uu=-1$,

-- \ for an absolute velocity  $\uu$, $\Sd_\uu:=\{\x\in\M\mid \uu\cdot\x=0\}$, the set of $\uu$-spacelike vectors,
is a three dimensional Euclidean vector space,

-- \ the action of linear and bilinear maps is denoted by a dot, too; 
e.g the action of a linear map $\Ll$ on a vector 
$\x$ is $\Ll\cdot\x$ (which is ${L^i}_kx^k$ in coordinates),

-- \ the adjoint of a linear map $\Ll$ is the linear map $\Ll^*$ defined by $(\Ll^*\cdot\x)\cdot\y=\x\cdot(\Ll\cdot\y)$ 
for all vectors $\x,\y$; shortly, $\Ll^*\cdot\x=\x\cdot\Ll$,

-- \  $\Ll$ is a Lorentz transformation if and only if $\Ll^*=\Ll^{-1}$,

-- \ the tensor product $\x\otimes\y$  of the vectors $\x$ and $\y$ is a linear map defined by 
$(\x\otimes\y)\cdot\z:=\x(\y\cdot\z)$ (in coordinates it is $x^iy^k$); $(x\otm\y)^*=\y\otm\x$; 
$\x\land\y:=\x\otm\y - (\x\otm\y)^*$ and $\x\lor\y:=\x\otm\y+(\x\otm\y)^*$ are the antisymmetric and symmetric tensor product,
respectively,

-- \ the formulae of electromagnetism are written in the choice $c=1$, $\hbar=1$ and the electric charge is measured by 
real numbers.

-- \ $\Dd$  denotes the differentiation in spacetime ($\pt_k$ in coordinates). 
The antisymmetric derivative of a vector valued function $\f$ is denoted by $\Dd\land\f$ 
($\pt_kf_i - \pt_if_k$ in coordinates). The spacetime divergence of $\f$ is denoted by $\Dd\cdot\f$ 
($\pt_kf^k$ in coordinates). 
Similarly,  $\Dd\cdot\T$ is the spacetime divergence of a tensor valued function $\T$ ($\pt_kT^{ik}$ in coordinates).

-- \ $\Dp$ denotes differentiation of other functions, e.g. parameterizations.
\medskip

The word `distribution' will appear in two different senses: 1. having its everyday meaning, 2. being a 
mathematical notion. To distinguish between them, I write distribution for the first one and Distribution for the second one.

The usual setting of Distribution Theory is based on $\rr^n$ (\cite{Gelf-Shil}, \cite{Demidov},\cite {Dijk}, \cite{Horvath}).
It is a quite simple generalization that   
spacetime and an observer space are taken instead of $\rr^4$ and $\rr^3$.
Another simple generalization is that vector and tensor Distributions are included, too.
The present article can be understood without a thorough knowledge of Distributions; besides the elementary notions the only non 
trivial one is {\bf pole taming}, described in the Appendix.

The guiding principle is that only quantities definable in Distribution Theory can make sense. 
Distributions are denoted by calligraphic letters. Locally integrable functions and the corresponding 
Distributions are distinguished in notation; e.g. $\F$ is an electromagnetic field function, $\Ff$ is 
the corresponding electromagnetic field Distribution.
\bigskip

\part*{Electrodynamics of point charges}

\section{The retarded proper time}

The existence of a material point in spacetime is described by a {\bf world line function} whose variable is the 
proper time; the range of a world line function is a {\bf world line}, a one dimensional submanifold in spacetime.

Let $r$ be a given continuously differentiable world line function. For a world point $x$,
$\s_r(x)$ will denote the {\it retarded proper time corresponding to $x$} i.e. the proper time point of $r$ 
for which $x-r(\s_r(x))$ is future lightlike or zero; evidently, it is zero if and only if 
$x$ is on the world line.  

In other words, $\s_r(x)$ is defined implicitly by the relations
$$ (x-r(\s_r(x)))\cdot(x-r(\s_r(x)))=0, \qquad -\uu\cdot(x-r(\s_r(x)))\ge0$$
for an arbitrary absolute velocity $\uu$.

According to the implicit function theorem, $\s_r$ is continuously differentiable outside the world line;
differentiating the first equality above, we obtain that
$2(x-r(\s_r(x))\cdot\bigl(\1 - \dot r(\s_r(x))\otm\Dd\s_r[x]\bigr)=0$ i.e.
$$ \Dd \s_r[x]=\frac{x-r(\s_r(x))}{\dot r(\s_r(x))\cdot(x-r(\s_r(x)))} \qquad (x\notin\ran r).$$

Though $\s_r$ is not differentiable on the world line, it is continuous there which can be seen as follows.

Let $\s_-<s_0<\s_+$ be arbitrary proper time points of the world line. If $T^\rightarrow$ denotes the set 
of futurelike vectors then 
$\bigl(r(\s_-) + T^\rightarrow\bigr)\cap\bigl(r(\s_+) - T^\rightarrow\bigr)$ is a bounded open set containing $r(\s_0)$. If $x$ 
is in this set then $r(\s_r(x))$ is in it, too; consequently the function $x\mapsto x-r(\s_r(x))$ is bounded. 

The proper time passed between two points of the world line is less or equal than the inertial time, thus
\begin{align*} |\s_r(x)-\s_0 |^2&\le -\bigl(r(\s_r(x)-r(\s_0))\bigr)\cdot\bigl(r(\s_r(x)-r(\s_0))\bigr) =\\
&=-\bigl(r(\s_r(x)-x) + (x -r(\s_0))\bigr)\cdot\bigl((r(\s_r(x)-x)+ (x -r(\s_0))\bigr)=\\
&= 2\bigl(x-r(\s_r(x))\bigr)\cdot(x-r(\s_0)) - (x-r(\s_0))\cdot(x-r(\s_0)).\end{align*}

Because of the mentioned boundedness, the right hand side tends to zero as $x$ tends to $r(\s_0)$.

\section{The electromagnetic field produced \\ by a point charge}

Let $r$ be a given twice differentiable world line function of a point charge $e$. 

Let us introduce the following functions:                                
\beq\label{retmeny} \R_r(x):= x - r(\s_r(x)), \qquad \uu_r(x):=\dot r(\s_r(x)), \qquad \ad_r(x):=\ddot r(\s_r(x)),\enq

\beq\label{lld} \Ll_r:=\frac{\R_r}{-\uu_r\cdot\R_r}=-\Dd\s_r, \qquad \dd_r:=\ad_r + (\ad_r\cdot\Ll_r)\uu_r.\enq

Then it is known (see \cite{Jackson}) that the electromagnetic field produced by a point charge $e$ with a
{\bf given world line function} $r$ is the regular Distribution $\Ff[r]$ defined by the locally integrable function
\beq\label{field} \F[r]:=\frac{e}{4\pi}\frac{\uu_r\land\Ll_r}{(-\uu_r\cdot\R_r)^2} + 
\frac{e}{4\pi}\frac{\dd_r\land\Ll_r}{-\uu_r\cdot\R_r};\enq
the first term is the {\it tied field} $\F[r]^{td}$ which is never zero, the second term is the 
{\it radiated field} $\F[r]^{rd}$ which is zero if and only if the charge is not accelerated.

It is emphasized by the notation $[r]$ that the formulae above make sense only if the world line function is {\bf given} 
because {\it the retarded proper time is defined only in that case}.

\section{`Energy-momentum tensor' of a point charge}\label{enmompont}

As usual, for an electromagnetic field function $\F$,
\beq\label{enimp}\T:=-\F\cdot\F - \frac1{4}\mathrm{Tr}\F\cdot\F\enq
is considered the energy-momentum tensor whose negative spacetime divergence $-\Dd\cdot\T$ is the spacetime force density.

The problem is that $\T$, in general, is not locally integrable, thus $\T$ does not define a Distribution a priori, so 
it has no physical meaning. In particular, this occurs for a point charge; that is why the quotation mark 
appeared in the section title. According to \eqref{field}, the `energy-momentum tensor' is

\beq\label{enimpt}\T[r]=\T[r]^{td} + \T[r]^{rtd} + \T[r]^{rd}\enq
where
$$\T[r]^{td}:=-\F[r]^{td}\cdot\F[r]^{td} - \frac1{4}\mathrm{Tr}\F[r]^{td}\cdot\F[r]^{td},$$
$$\T[r]^{rtd}:=-(\F[r]^{td}\cdot\F[r]^{rd}+\F[r]^{rd}\cdot\F[r]^{td}) - 
\frac1{4}\mathrm{Tr}(\F[r]^{td}\cdot\F[r]^{rd}+\F[r]^{rd}\cdot\F[r]^{td}),$$
$$\T[r]^{rd}:=-\F[r]^{rd}\cdot\F[r]^{rd} - \frac1{4}\mathrm{Tr}\F[r]^{rd}\cdot\F[r]^{rd},$$

for which we have the following equalities:

\beq\label{enm1} -\F^{td}[r]\cdot\F^{td}[r]=\frac{e^2}{16\pi^2}\frac{\uu_r\otimes\Ll_r + \Ll_r\otimes\uu_r - \Ll_r\otimes\Ll_r}
{(\uu_r\cdot\R_r)^4},\enq

\beq\label{enm2}-\mathrm{Tr}(\F^{td}[r]\cdot\F^{td}[r])=\frac{e^2}{16\pi^2}\frac2{(\uu_r\cdot\R_r)^4},\enq

\beq\label{enm3} -\bigl(\F^{td}[r]\cdot\F^{rd}[r] + \F^{rd}[r]\cdot\F^{td}[r]\bigr)=
\frac{e^2}{16\pi^2}\frac{\dd_r\otimes\Ll_r + \Ll_r\otimes\dd_r +
2(\uu_r\cdot\dd_r)\Ll_r\otimes\Ll_r}{(-\uu_r\cdot\R_r)^3},\enq

\beq\label{enm4}\mathrm{Tr}(\F^{td}[r]\cdot\F^{rd}[r])=0, \enq

\beq\label{enm5} -\F^{rd}[r]\cdot \F^{rd}[r]=\frac{e^2}{16\pi^2}\frac{|\dd_r|^2\Ll_r\otimes\Ll_r}{(\uu_r\cdot\R_r)^2},\enq

\beq\label{enm6}\mathrm{Tr}(\F^{rd}[r]\cdot\F^{rd}[r])=0.\enq
      
Evidently, $\T[r]$ is differentiable outside the world line; its properties near the world line will be examined in
the forthcoming sections.

\section{Radial expansion of the retarded proper time}

The retarded proper time is not differentiable on the world line, so it has not a Taylor polynomial around the world line.
Nevertheless, it can be expanded in `radial directions' in powers of the `radial distance' as follows.

We take the parameterization \eqref{param} of spacetime around a given world line $\ran r$; in this parameterization,
at every proper time point $\s$ of $r$, the spacelike vectors Lorentz orthogonal to $\dot r(\s)$ are taken into account; 
then according to the notations 
introduced in \eqref{rno}, $\rho$ and $\nb$ describe distances and directions, respectively, in 
that Euclidean space. The difference between the parameter proper time and the retarded proper time is
       
$$\Theta(\s,\q):=\s - \s_r\bigl(r(s) +\Ll(\s)\q\bigr).$$

$\Theta$ is a continuous function,  non-negative and  $\Theta(\s,\q)=0$ if and only if $\q=0$. Thus,  
\beq\label{limth}\lim_{\q\to\0}\Theta(\s,\q)=0\enq
holds as well.

In the sequel, $r$ is supposed to be three times continuously differentiable, the variables are omitted for 
the sake of brevity and $\Ord$ denotes a function whose limit at zero is zero.  

First, let us observe that according to \eqref{limth} 
$$ \Theta=\Ord(\rho).$$

Then starting with $\s_r=\s-\Theta$, we have
\beq\label{rcircr} r(\s -\Theta)= r(\s) - \dot r(\s)\Theta + \frac1{2}\ddot r(\s)\Theta^2 - \frac1{6}\dddot r(\s)\Theta^3 + 
\Theta^3\Ord(\Theta).\enq

Subtracting both sides from the parameterization \eqref{rron}, we get (see \eqref{retmeny})

\beq\label{Rr} \R_r=\rho\nb + \dot r\Theta - \frac1{2}\ddot r\Theta^2 + \frac1{6}\dddot r\Theta^3 + 
\Theta^3\Ord(\Theta).\enq

The left hand side is a lightlike vector or zero. The Lorentz square of both sides, with the equalities
$$ \dot r\cdot\dot r=-1,\ \ \nb\cdot\nb=1, \ \ \dot r\cdot\nb=0, \ \ \dot r\cdot\ddot r=0, \ \
\dot r\cdot\dddot r=-\ddot r\cdot\ddot r,$$
results in
$$ 0=\rho^2 - (1+\rho\nb\cdot\ddot r)\Theta^2 + \frac1{3}\rho\nb\cdot\dddot r\Theta^3 
-\frac1{12}\ddot r\cdot\ddot r\Theta^4 +
(\rho\Theta^3 +\Theta^4)\Ord(\Theta).$$

In another form,
\beq\label{teta}1=\frac{\Theta^2}{\rho^2}\left(1+\rho\nb\cdot\ddot r - \frac1{3}\rho\nb\cdot\dddot r\Theta +
\frac1{12}|\ddot r|^2\Theta^2 - (\rho\Theta+\Theta^2)\Ord(\Theta)\right),\enq
which says that $\lim_{\rho\to0}\frac{\Theta^2}{\rho^2}=1$. Since $\Theta$ is non-negative, 
$\lim_{\rho\to0}\frac{\Theta}{\rho}=1$ holds as well. Then
\beq\label{teta0}\Theta=\rho(1+ A) \qquad \text{where} \ A=\ord\enq
and \eqref{teta} can be rewritten in the form
$$1=\frac{\Theta^2}{\rho^2} + \frac{\Theta^2}{\rho^2}(\rho\nb\cdot\ddot r + \rho\ord).$$

Putting \eqref{teta0} in the first term on the right hand side, we get
$$\frac{2A + A^2}{\rho}= -\frac{\Theta^2}{\rho^2}(\nb\cdot\ddot r +\ord).$$
Let $\rho$ tend to zero. Since $A=\ord$, we get
$$ \lim_{\rho\to}\frac{2A}{\rho}=-\nb\cdot\ddot r,$$
in other words, $A=\rho(-\frac1{2}\nb\cdot\ddot r + \ord)$; thus,
\beq\label{teta1} \Theta = \rho -\frac{\rho^2}{2}(\nb\cdot\ddot r + B) \qquad \text{where} \ B=\ord.\enq

Then \eqref{teta} can be written in the form:
\beq\label{utolso}1=\frac{\Theta^2}{\rho^2}(1+\rho\nb\cdot\ddot r) -
\frac{\Theta^2}{\rho^2}\left(\frac1{3}\rho\nb\cdot\dddot r\Theta -
\frac1{12}|\ddot r|^2\Theta^2 + \rho^2\ord\right).\enq

Putting \eqref{teta1} in the first term on the right hand side, we get
\begin{multline*}\left(1 - \rho\nb\cdot\ddot r - \rho B +\frac{\rho^2}{4}\bigl((\nb\cdot\ddot r)^2 + 
2\nb\cdot\ddot r B + B^2\bigr)\right)(1 + \rho\nb\cdot\ddot r)= \\
=1 -\rho^2(\nb\cdot\ddot r)^2 - \rho B + \frac{\rho^2}{4}\bigl((\nb\cdot\ddot r)^2 + \ord\bigr).\end{multline*}
Then \eqref{utolso} divided by $\rho^2$ becomes 
$$ 0= -(\nb\cdot\ddot r)^2 - \frac{B}{\rho} + \frac{1}{4}\bigl((\nb\cdot\ddot r)^2 + \ord\bigr)
-\frac{\Theta^2}{\rho^2}\left(\frac1{3}\nb\cdot\dddot r\frac{\Theta}{\rho} - 
\frac1{12}|\ddot r|^2\frac{\Theta^2}{\rho^2} + \ord\right)$$ 
from which
$$ \lim_{\rho\to0}\frac{B}{\rho}=-\frac3{4}(\nb\cdot\ddot r)^2 -\frac1{3}\nb\cdot\dddot r +\frac1{12}|\ddot r|^2.$$
Then
$$ B=\rho\left(-\frac3{4}(\nb\cdot\ddot r)^2 -\frac1{3}\nb\cdot\dddot r+\frac1{12}|\ddot r|^2 + \ord)\right).$$

Then we have the final result

\beq\label{teta2} \Theta=\rho\left(1 - \rho\frac{\nb\cdot\ddot r}{2} + 
\frac{\rho^2}{2}\left(\frac{9(\nb\cdot\ddot r)^2 
- |\ddot r|^2}{12}\ + \frac{\nb\cdot\dddot r}{3}\right) + \rho^2\ord\right).\enq

For the sake of simplicity, without the danger of confusion, we shall omit the terms $\ord$, writing 
approximative equalities:

\beq\label{thapp}\Theta\app\rho\left(1 - \rho\frac{\nb\cdot\ddot r}{2} + 
\frac{\rho^2}{2}\left(\frac{9(\nb\cdot\ddot r)^2 
- |\ddot r|^2}{12}\ + \frac{\nb\cdot\dddot r}{3}\right)\right),\enq 

\beq\label{thapp1}\Theta^2\app\rho^2(1-\rho(\nb\cdot\ddot r)),\qquad \Theta^3\app\rho^3.\enq

It is emphasized again that these formulae do not come from a Taylor expansion because of the non differentiability 
of the retarded proper time on the world line. 

\section{Further radial expansions}

Using the previous results, we give the radial expansion of the functions occurring in the formulae 
of the electromagnetic field \eqref{field}.

According to \eqref{thapp} and \eqref{thapp1}, the retarded velocity and acceleration introduced in \eqref{retmeny} have radial
expansion
\beq\label{urp} \uu_r \app \dot r - \ddot r\Theta + \frac1{2}\dddot r\Theta^2 \app
\dot r -\rho\ddot r + \frac{\rho^2}{2}\bigl((\nb\cdot\ddot r)\dot r + \dddot r\bigr),\enq
\beq \ad_r\app \ddot r - \dddot r\Theta \app\ddot r -\rho\dddot r.\enq
        
Further, \eqref{Rr} gives us
\begin{align*}\R_r\app& \rho\nb + \dot r\Theta - \frac1{2}\ddot r\Theta^2 + \frac1{6}\dddot r\Theta^3 \app \\
\app& \rho\Biggl(\nb + \ddot r -\rho\frac{\ddot r +(\nb\cdot\ddot r)\dot r}{2} +\\
& \qquad \ + \frac{\rho^2}{2}\left(\left(\frac{9(\nb\cdot\ddot r)^2 
- |\ddot r|^2}{12} + \frac{\nb\cdot\dddot r}{3}\right)\dot r + (\nb\cdot\dot r)\ddot r + \frac{\dddot r}{3}\right)\Biggl), \end{align*}

$$(-\uu_r\cdot\R_r)\app\rho\left(1 +\rho\frac{(\nb\cdot\ddot r)}{2} - 
\frac{\rho^2}{2}\left(\frac{(\nb\cdot\ddot r)^2 -|\ddot r|^2}{4}+\frac{2\nb\cdot\dddot r}{3}\right)\right), $$

$$\frac1{-\uu_r\cdot\R_r} \app\frac1{\rho}\Biggl(1-\rho\frac{\nb\cdot\ddot r}{2} 
+\frac{\rho^2}{2}\left(\frac{3(\nb\cdot\ddot r)^2 -|\ddot r|^2}{4} + \frac{2\nb\cdot\dddot r}{3}\right)\Biggr),$$

$$\frac1{(\uu_r\cdot\R_r)^2}\app\frac1{\rho^2}\Biggl(1-\rho\nb\cdot\ddot r +
\frac{\rho^2}{2}\left(\frac{4(\nb\cdot\ddot r)^2 -|\ddot r|^2}{2} + \frac{4\nb\cdot\dddot r}{3}\right)\Biggr),$$

$$\label{uR3}\frac1{(-\uu_r\cdot\R_r)^3} \app\frac1{\rho^3}\Biggl(1-\rho\frac{3\nb\cdot\ddot r}{2} 
+\frac{\rho^2}{2}\left(\frac{15(\nb\cdot\ddot r)^2 -3|\ddot r|^2}{4} + 2\nb\cdot\dddot r\right)\Biggr),$$

$$\label{uR4}\frac1{(\uu_r\cdot\R_r)^4}\app\frac1{\rho^4}\Biggl(1-\rho2\nb\cdot\ddot r 
+\frac{\rho^2}{2}\left(6(\nb\cdot\ddot r)^2 -|\ddot r|^2 + \frac8{3}\nb\cdot\dddot r\right)\Biggr).$$

Thus, for the quantities introduced in \eqref{lld},
\beq\label{lrp}\begin{split}\Ll_r\app \nb+\dot r -&\rho\left(\frac{\ddot r}{2} + (\nb\cdot\ddot r)\left(\dot r + 
\frac{\nb}{2}\right)\right)+\\ 
&+\frac{\rho^2}{2}\left(\frac{6(\nb\cdot\ddot r)^2 -|\ddot r|^2}{3} +\nb\cdot\dddot r\right)\dot r + \\
&+\frac{\rho^2}{2}\left(\frac{3(\nb\cdot\ddot r)^2 - |\ddot r|^2}{4} + \frac{2\nb\cdot\dddot r}{3}\right)\nb +\\
&+\frac{\rho^2}{2}\left(\frac{3(\nb\cdot\ddot r)\ddot r}{2} + \frac{\dddot r}{3}\right),\end{split}\enq

\beq\label{drp} \dd_r\app\ddot r + (\nb\cdot\ddot r)\dot r + \rho\left(\left(\frac{|\ddot r|^2 - (\nb\cdot\ddot r)^2}{2} - 
\nb\cdot\dddot r\right)\dot r - (\nb\cdot\ddot r)\ddot r -\dddot r\right)\enq

\section{Radial expansion of the \\ `energy-momentum tensor'}\label{radexpmom}

Let us take \eqref{enm1}-\eqref{enm6} and, for the sake of simplicity, let us rewrite \eqref{lrp} in the form 
$$ \Ll_r\app \nb + \dot r -\rho \A +\frac{\rho^2}{2}\B.$$
Then -- $\lor$ denotes the symmetric tensor product  --
\begin{align*} (\uu_r\lor\Ll_r)&\app\uu\lor(\nb +\uu) -\rho\bigl(\uu\lor\A +\ddot r\lor(\nb+\uu)\bigr) +\\
&+\frac{\rho^2}{2}\Bigl(\dot r\lor\B +((\nb\cdot\ddot r)\ddot r +\dddot r)\lor(\nb+\dot r) +2\ddot r\lor\A\Bigr),\end{align*}
\begin{align*} (\Ll_r\otm\Ll_r)&\app(\nb+\dot r)\otm(\nb+\dot r) - \rho\big((\nb+\dot r)\lor\ddot r\bigr) +\\
&+\frac{\rho^2}{2}\Bigl((\nb +\dot r)\lor\B +\A\lor\A\Bigr).\end{align*}

Further, rewriting \eqref{drp} in the form
$$ \dd_r\app \ddot r +(\nb\cdot\ddot r)\dot r + \rho(\beta\dot r - (\nb\cdot\ddot r)\ddot r - \dddot r)$$
we have
\begin{align*} (\dd_r\lor\Ll_r)&\app (\ddot r +(\nb\cdot\ddot r)\dot r)\lor(\nb+\dot r) + \\ 
&+\rho\Bigl(-\bigl(\ddot r +(\nb\cdot\ddot r)\dot r\bigr)\lor\A + (\beta\dot r-(\nb\cdot\ddot r)\ddot r 
-\dddot r)\lor(\nb+\dot r)\Bigr),\end{align*}
\beq (\uu_r\cdot\dd_r)\app -(\nb\cdot\ddot r +\rho\beta),\enq
and 
\begin{align*}(\uu_r\cdot\dd_r)(\Ll_r\otm\Ll_r)&\app -(\nb\cdot\ddot r)\bigl((\nb+\dot r)\otm(\nb+\dot r)\bigr) +\\
&+\rho\bigl((\nb\cdot\ddot r)(\nb+\dot r)\lor\A - \beta(\nb+\dot r)\otm(\nb+\dot r)\bigr);\end{align*}
finally,
$$ |\dd_r|^2\app |\ddot r|^2 - (\nb\cdot\ddot r)^2.$$

All these together give that the `energy-momentum tensor' \eqref{enimpt} has the radial expansion 
\beq\label{Tkifejt}\begin{split}
\frac{16\pi^2}{e^2}\T[r] &= \frac{\dot r\otm\dot r - \nb\otm\nb +\frac{\1}{2}}{\rho^4} - \\
&+\frac{\frac{\nb\otm\ddot r + \ddot r\otm\nb}{2} -(\nb\cdot\ddot r)(2\dot r\otm\dot r - \nb\otm\nb +\1)}{\rho^3} -\\ 
&-\frac{1}{\rho^2}\left(\frac{\ddot r\otimes\ddot r}{4} + \frac{\dot r\otimes\dddot r + \dddot r\otimes\dot r}{2} \right) + \\
&-\frac1{\rho^2}\left(\frac{2(\nb\otimes\dddot r + \dddot r\otimes\nb)}{3} + 
(\nb\cdot\ddot r)(\nb\otimes\ddot r + \ddot r\otimes\nb)\right) +\\
&+\frac1{\rho^2}\left(|\ddot r|^2 + 3(\nb\cdot\ddot r)^2 + \frac{4\nb\cdot\dddot r}{3}\right)\dot r\otimes\dot r + \\
&+\frac1{\rho^2}\left(\frac{3|\ddot r|^2}{4} -(\nb\cdot\ddot r)^2\right)\nb\otimes\nb + \\
&+\frac1{\rho^2}\left(\frac{2|\ddot r|^2}{3} +\frac{\nb\cdot\dddot r}{2}\right)(\dot r\otimes\nb + \nb\otimes\dot r) + \\
&+\frac1{\rho^2}\left(\frac{|\ddot r|^2}{4} + \frac{3(\nb\cdot\ddot r)^2}{2} + \frac{2\nb\cdot\dddot r}{3}\right)\1 + 
\frac{\ord}{\rho^2}.\end{split}\enq

\section{Distribution of energy-momentum}

The radial expansion \eqref{Tkifejt} shows that the `energy-momentum tensor' $\T[r]$, for every proper time value $\s$ of the
given world line function $r$,  has a pole of fourth order and a pole of third order in the three dimensional Euclidean 
space Lorentz orthogonal to
$\dot r(\s)$.  We can tame its poles, obtaining $\tm\T[r]$ as follows. For the sake of avoiding misunderstandings, we shall 
write simply $\T$ instead of $\T[r]$ when integrating functions.

For an $x$ in the neighbourhood of the world line in question let $\hat\s(x)$ denote the proper time value
for which 
\beq\label{hats} \dot r(\hat\s(x))\cdot\bigl(x-r(\hat\s(x)\bigr)=0,\enq
holds, i.e. $\hat\s$ is the first component of the inverse of the parameterization \eqref{param}.

\begin{all} If the support of the test function $\phi$ is disjoint from the range of $r$ then 
$$ \bigl(\tm\T[r]\mid\phi\bigr):=\intl_\aM \T(x)\phi(x)\ dx;$$
if the support of the test function $\phi$ is in a world roll $H_{I,R_I}$ then
\begin{multline}\label{depolm}
 \bigl(\tm\T[r]\mid\phi\bigr):=\\ :=\intl_{\aM}\Biggl(\T(x)\phi(x)  
-\frac{e^2}{16\pi^2}\frac{4\dot r(\hat\s(x))\otimes\dot r(\hat\s(x)) +\1}
{6|x- r(\hat\s(x))|^4}\phi(r(\hat\s(x)))\Biggr)\ dx \ !!
\end{multline} 
\noindent where the double exclamation mark says the integral must be taken in the radial parameterization of the world roll 
$H_{I,R_I}$ and in a given order.\end{all}

{\bf Proof} The integral \eqref{depolm} in the radial parameterization reads
\begin{align*} \intl_\I\Biggl(\intl_0^\infty\Biggl(\intl_{S_1(0)}&\Bigl(\T(r(\s)+\rho\nb)\phi(r(s) + \rho\nb) -\\
&-\frac{e^2}{16\pi^2}\frac{4\dot r(\s)\otimes\dot r(\s) +\1}{6\rho^4}\phi(r(\s))\Bigr) 
(1 + \rho(\nb\cdot\ddot r(\s))\ d\nb_0\Biggr)\ \rho^2 d\rho\Biggr)  \ d\s\end{align*}
with $\nb=\Ll(\s)\nb_0$.

Let us write the radial expansion of the `energy-momentum tensor' in the form
\beq \frac{A(\nb)}{\rho^4} + \frac{B(\nb)}{\rho^3} + \frac{C(\nb) + \Ord(\rho)}{\rho^2}.\enq

Then the function
\begin{align} &\left(\frac{A(\nb)}{\rho^4} + \frac{B(\nb)}{\rho^3} + \frac{C(\nb) + \Ord(\rho)}{\rho^2}\right)
(1 + \rho\nb\cdot\ddot r)=\\
&=\label{ttomp}\frac{A(\nb)}{\rho^4} + \frac{A(\nb)(\nb\cdot\ddot r) + B(\nb)}{\rho^3} + 
\frac{B(\nb)(\nb\cdot\ddot r) + C(\nb) + \ord}{\rho^2} \end{align}
is to be pole tamed for each fixed $\s$. 

$A(\nb)$ is an even tensor product of $\nb$ (linear function of an even tensor muliple of $\nb_0$),
so the first term of \eqref{ttomp} can be tamed: the function
$$ \frac{A(\nb)\bigl(\phi(r(\s) + \rho\nb)- \phi(r(\s))\bigr)}{\rho^4}$$
is to be integrated in the radial variables (because the derivative of $\phi$ drops out from the Taylor polynomial).

The part containing $\phi(r(\s) + \rho\nb)$ remains as it is,
\beq\label{tomp1} \frac{A(\nb)}{\rho^4}\phi(r(\s) + \rho\nb);\enq
the part containing $\phi(r(\s))$ becomes
\beq\label{tompk} \frac{e^2}{16\pi^2}\left(\frac{4\dot r(s)\otm\dot r(\s) + \1}{6\rho^4}\right)\phi(r(\s))\enq
which is obtained from the integral
$$\frac1{4\pi}\int_{S_1(0)}\left(\dot r(\s)\otm\dot r(\s) - \Ll(\s)\nb_0\otm\Ll(\s)\nb_0 + \frac{\1}{2}\right)\ d\nb_0$$
by the use of
$$ \intl_{S_1(\0)}\Ll(\s)\nb_0\otm\Ll(\s)\nb_0 \ d\nb_0=
\frac{4\pi}{3}\bigl(\1 + \Ll(\s)\dot r(0)\otm\Ll(\s)\dot r(0)\bigr)=
\frac{4\pi}{3}\bigl(\1 + \dot r(\s)\otm\dot r(\s)\bigr).$$

$A(\nb)\nb\cdot\ddot r + B(\nb)$ is a linear function of an odd tensor product of $\nb$, so the second term in \eqref{ttomp} 
can be tamed: the function
\beq\label{tomp2} \frac{A(\nb)\nb\cdot\ddot r + B(\nb)}{\rho^3}\phi(r(s) + \rho\nb)\enq
is to be integrated in the radial variables (because $\phi(r(\s))$ drops out from the Taylor polynomial).

The third term in \eqref{ttomp} defines a regular Distribution, so the function 
\beq\label{tomp3} \frac{B(\nb)\nb\cdot\ddot r + C(\nb) + \ord}{\rho^2}\phi(r(\s) +\rho\nb)\enq 
is to be integrated.

Summarizing: the sum of \eqref{tomp1}, \eqref{tomp2} and \eqref{tomp3} gives the first term in \eqref{depolm} 
and \eqref{tompk} gives the second term because in the radial parameterization $\hat\s(x)=\s$ and $|x -r(\hat\s(x))|=\rho$. $\mqed$

$\tm\T[r]$ will be called the {\it Distribution of energy-momentum} (which is not an energy-momentum distribution!) 
of the point charge with a given world line function $r$. 

\section{The radiation reaction force}

The Distribution of energy-momentum of a point charge is a mathematical object. Has it some physical meaning? 
The negative spacetime divergence of the energy-momentum tensor of a continuous charge-current density is the force density; 
that is why we can hope here a similar physical meaning.

Recall that $\la_{\ran r}$ is the Lebesgue measure of the world line $\ran r$ (see \eqref{ranr}).

\begin{all} 
\begin{align} -\Dd\cdot\tm\T[r] & 
=\frac{1}{4\pi}\frac{2e^2}{3} \bigl(\dddot r + (\dot r\cdot\dddot r)\dot r\bigr)\la_{\ran r} =\\
&= \label{visszero}\frac{1}{4\pi}\frac{2e^2}{3}(\dot r\land\dddot r)\cdot\dot r \la_{\ran r}. \end{align}\end{all}
\sk
 
{\bf Proof} The  Distribution $\tm\T[r]$ restricted to an open subset disjoint from the world line 
equals the regular Distribution corresponding to the restriction of $\T[r]$. Since  
$\Dd\cdot\T[r]=0$ except the world line, we have to take test functions 
$\phi$ whose support is in a world roll $H_{I,R_I}$ around the world line. For them
\begin{align*} &(-\Dd\cdot\tm\T[r])\mid\phi)=(\tm\T[r]|\cdot\Dd\phi)=\\ 
&= \intl_\aM\Biggl(\T(x)\cdot\Dd\phi[x]  
-\frac{e^2}{16\pi^2}\frac{4\dot r(\hat\s(x))\otm\dot r(\hat\s(x)) +\1}
{6|x-r(\hat\s(x)|^4}\cdot\Dd\phi[r(\hat\s(x))]\Biggr)\ dx \ !! =\\
&\qquad\qquad= \lim_{R\to0}\intl_{\aM\setminus H_{I,R}} \dots \ !!\end{align*}

Because of $\Dd\cdot\T=0$ in $\aM\setminus H_{I,R}$,, the first term in the integrand equals $\Dd\cdot(\T\phi)$.

For examining the second term let us consider $\rho$ and $\nb$ as functions of the spacetime points
$x$ instead of the radial parameters $(\s,\q)$, i.e. let us introduce
\beq \label{rokalap}\hat\rho(x):=|x-r(\hat\s(x))|, \quad \hat\nb(x):=\frac{x- r(\hat\s(x))}{|x-r(\hat\s(x))|};\enq
moreover, let
$$ \hat\uu(x):=\dot r(\hat\s(x)), \qquad \hat\ad(x):=\ddot r(\hat\s(x)).$$

Then
$$ (\hat\rho\hat\nb)(x)=x-r(\hat\s(x))$$
and
\beq\label{nbunb} \hat\nb\cdot\hat\nb=1, \qquad \hat\uu\cdot\hat\nb=0.\enq

By differentiation we get
\beq\label{rondif} \1 - \hat\uu\otm\Dd\hat\s =\Dd(\hat\rho\hat\nb)=\hat\nb\otm\Dd\hat\rho + \hat\rho\Dd\hat\nb, \enq
from which, by \eqref{nbunb} we deduce
\beq\label{dro} \Dd\hat\rho=\hat\nb.\enq
Further,
$$ \Dd\hat\uu =\hat\ad\otm\Dd\hat\s,$$
and 
$$ 0=\Dd(\hat\uu\cdot\hat\rho\hat\nb)=\hat\rho\hat\nb\cdot(\hat\ad\otm\Dd\hat\s) + \hat\uu\cdot(\1 - \hat\uu\otm\Dd\hat\s),$$
therefore
\beq\label{dhats}\Dd\hat\s= -\frac{\hat\uu}{1 + \hat\rho\hat\nb\cdot\hat\ad}\enq
and as a consequence,
\beq\label{divron}\Dd\cdot(\hat\rho\hat\nb)=4 - \frac1{1 + \hat\rho\hat\nb\cdot\hat\ad},\enq

Further, let  

$$ \dZ(\s):=\frac{(4\dot r(\s)\otimes\dot r(\s) + \1)\cdot\Dd\phi[r(\s)]}{6}, \qquad \hat\dZ(x):= \dZ(\hat\s(x)).$$

Then we find that 
$\Dd\hat\dZ=\dot\dZ(\hat\s)\otm\Dd\hat\s$ from which by \eqref{dhats} $(\Dd\hat\dZ)\cdot\hat\nb=0$ follows.
Then taking into account \eqref{divron} and \eqref{dro},
\begin{align*}\Dd\cdot\frac{\hat\dZ\otm\hat\rho\hat\nb}{\hat\rho^4}=
& \hat\dZ\left(\left(4 - 
\frac1{1 + \hat\rho\hat\nb\cdot\hat\ad}\right)\frac1{\hat\rho^4} -\frac{\hat\rho\hat\nb\cdot 4\hat\nb}{\hat\rho^5}\right) = \\
=&- \frac{\hat\dZ}{(1 + \hat\rho\hat\nb\cdot\hat\ad)\hat\rho^4}=-\frac{\hat\dZ}{\hat\rho^4} 
+ \frac{\hat\dZ(\hat\nb\cdot\hat\ad)}{(1 + \hat\rho\hat\nb\cdot\hat\ad)\hat\rho^3},\end{align*}
in other words, 
$$ -\frac{\hat\dZ}{\hat\rho^4} = \Dd\cdot\frac{\hat\dZ\otm\hat\nb}{\hat\rho^3}-
 \frac{\hat\dZ(\hat\nb\cdot\hat\ad)}{(1 + \hat\rho\hat\nb\cdot\hat\ad)\hat\rho^3}.$$

Thus, we arrive at
\begin{multline*}(-\Dd\cdot\tm\T[r]\mid\phi)= \\
=\lim_{R\to0}\intl_{\aM\setminus H_{I,R}}\Biggl(\Dd\cdot\Biggl(\T\phi +  
\frac{e^2}{16\pi^2}\frac{\hat{\dZ}\otimes\hat\nb}{\hat\rho^3}\Biggr)  
-\frac{e^2}{16\pi^2}\frac{\hat{\dZ}(\nb\cdot\hat\ad)}
{(1 + \hat\rho\hat\nb\cdot\hat\ad)\hat\rho^3}\Biggr).\ !!\end{multline*}
      
The integral of the second term in the parameterization around the world line becomes a multiple of
$$ \intl_I\intl_R^{R_I}\rho^2\intl_{S_1(\0)}\frac{\dZ(\s)(\Ll(\s)\cdot\nb_0)\cdot\ddot r(\s)}{\rho^3}\ d\nb_0\ d\rho\ d\s$$
which is zero because of \eqref{n1}.

By Gauss' theorem, the integral of the first term can be transformed to an integral on the 
corresponding world tube in such a way that the tube is directed `inwards' i.e. the normal vector is $-\hat\nb$:
\beq\begin{split} (-\Dd\cdot\tm\T[r]\mid\phi) &=
\lim_{R\to0}\intl_{\aM\setminus H_{I,R}}\Dd\cdot\Biggl(\T\phi +  
\frac{e^2}{16\pi^2}\frac{\hat{\dZ}\otimes\hat\nb}{\hat\rho^3}\Biggr)\ !! =\\
&=- \lim_{R\to0}\intl_{P_{I,R}}\Biggl(\T\phi +\frac{e^2}{16\pi^2}
\frac{\hat{\dZ}\otimes\hat\nb}{\hat\rho^3}\Biggr)\cdot\hat\nb\ d\la_{P_{I,R}}\ !!= \\
&=-\label{kifint} \lim_{R\to0}\intl_IR^2\intl_{S_1(0)} ...(\nb_0,\s)\ d\nb_0\ d\s \ !\end{split}\enq
where the integrand is
\beq\label{kifintt} \Biggl((\T\phi)\bigl(r(\s) + R\nb(\s)\bigr) + 
\frac{e^2}{16\pi^2}\frac{\dZ(\s)\otimes\nb(\s)}{R^3}\Biggr)\cdot\nb(\s)
(1 + R\nb(\s)\cdot\ddot r(\s))\enq 
whith  $\nb(\s):=\Ll(\s)\nb_0$. 

Then 
$$\frac{\dZ\otm\nb}{\R^3}\cdot\nb(1+R\nb\cdot\ddot r)= \frac{(4\dot r\otm\dot r +\1)\cdot\Dd\phi[r]}{6}
\left(\frac1{R^3} + \frac{\nb\cdot\ddot r}{R^2}\right),\ $$
and on the base of the radial expansion \eqref{Tkifejt}  
\begin{align*}(\T(r +R\nb)&\cdot\nb(1+R\nb\cdot\ddot r)= \\ 
&=\frac{e^2}{16\pi^2}\Bigl(-\frac{\nb}{2R^4} + \frac{\dot r}{2R^3}+ \\ &+ \frac{|\dot r|^2\nb + \frac2{3}|\dot r|^2\dot r 
-\frac3{4}(\nb\cdot\dot r)\dot r - \frac2{3}\dddot r}{R^2} +\\ &+ \frac{\Ord(R)}{R^2}\Bigr).\end{align*}

This, together with the expansion    
$$\phi(r + R\nb) = \phi(r)+ R\nb\cdot\Dd\phi[r] + R^2\Ord(R),$$
give that the integral \eqref{kifint} of the Ordo terms in \eqref{kifintt}, because of the  three times differentiability of $r$, 
is less than a multiple of $R$, so their limit is zero when $R$ tends to zero; moreover, the integrals of the terms linear and 
trilinear in $\nb$ in \eqref{kifintt} are zero.

The integral of the terms independent of $\nb$  results in a multiplication by
$4\pi $, the integral of the terms bilinear in $\nb$ results in a multiplication by
$\frac{4\pi}{3}(\1 + \dot r\otm\dot r)$. 

Thus, $\Dd\phi[r]$ will be multiplied by 
$$-\frac{4\pi}{3}\frac{\1 + \dot r\otimes\dot r}{2R} + 4\pi\frac{4\dot r\otimes\dot r + \1}{6R}=
\frac{4\pi}{2R}\dot r\otimes\dot r$$
and $\phi(r)$ will be multiplied by $\frac{4\pi}{2R}\ddot r + \frac2{3}\left(|\ddot r|^2\dot r - \dddot r\right)$. 
Since 
\beq\label{ints}\frac{4\pi}{2R}\Bigl(\dot r(\s)(\dot r(\s)\cdot\Dd\phi[r(\s)]) + \frac{4\pi}{2R}\ddot r(\s)\Bigr)= 
\frac{4\pi}{2R}\frac{d}{d\s}\bigl(\dot r(\s)\phi(r(\s))\bigr)\enq
and its integral is zero because the support of $\phi$ is in $H_{I,R_I}$, it remains only 
$$-\frac{e^2}{16\pi^2}\intl_I 4\pi\frac2{3}(|\ddot r(\s)|^2\dot r(\s) - \dddot r(\s) )\phi(r(\s)) \ d\s$$
which gives the desired result by $|\ddot r|^2=-\dot r\cdot\dddot r$. $\mqed$ 

\section{Discussion}

\subsection{On the self-force} 

The usual formula \eqref{visszero} of the radiation reaction force is obtained in a mathematically exact way, 
without unjustified application of Gauss--Stokes theorem, of Taylor expansion and without a doubtful limit to zero. 

It is an important fact that the self-force is obtained for a {\bf given} world line function $r$. 
Consequently, its physical meaning is the following: 
\medskip

{\it If the world line function is $r$ then 
the self-force  is $\frac{1}{4\pi}\frac{2e^2}{3}(\dot r\land\dddot r)\cdot\dot r$.}
\medskip

Accordingly, if the force $\f$ is necessary to get a given $r$ 
without radiation i.e. $m\ddot r=\f(r,\dot r)$ would valid without radiation, then besides $\f$ the force 
opposite to the self-force must be applied for getting the desired world line function $r$.

An actual example is when an elementary particle is revolved in a cyclotron by a homogeneous static 
magnetic field. In spacetime formulation: there is given a constant electromagnetic field $\F$
and the world line without radiation would satisfy $m\ddot r=e\F\dot r$. Then
$$ \dddot r=\frac{e^2}{m^2}\F\cdot\F\cdot\dot r, \qquad \dot r\cdot\dddot r= 
\frac{e^2}{m^2}(\dot r\cdot\F)\cdot(\F\cdot\dot r) = - \frac{e^2}{m^2}(\F\cdot\dot r)\cdot(\F\cdot\dot r),$$
thus the self-force is
\beq\label{komp}\frac1{4\pi}\frac{2e^4}{3m^2}\bigl(\F\cdot\F\cdot\dot r -|\F\cdot\dot r|^2\dot r\bigr);\enq
to keep the particle on a prescribed orbit, the opposite of the above force must be applied 
besides the magnetic field,

For better seeing, let us take the standard inertial frame $\uu$ in which the cyclotron is at rest. Then the 
$\uu$-spacelike component of $\F$, for which $\F\cdot\uu=0$, is the magnetic field $\underline\B$, 
a three dimensional antisymmetric tensor.

Using the relative velocity $\vv:=\frac{\dot r}{-\uu\cdot\dot r} - \uu$ with respect to the inertial frame, we have
$$ \F\cdot\dot r= \F\cdot\left(\frac{\dot r}{-\uu\cdot\dot r} - \uu\right)(-\uu\cdot\dot r)=
\frac{\underline\B\cdot\vv}{\sqrt{1-|\vv|^2}}$$
and the self-force relative to the inertial frame is  
\begin{equation*}\frac1{4\pi}\frac{2e^4}{3m^2}\frac{(\1 + \uu\otm\uu)}{-\uu\cdot\dot r}
\bigl(\F\cdot\F\cdot r -|\F\cdot\dot r|^2\dot r\bigr)=\frac1{4\pi}\frac{2e^4}{3m^2}\left(\underline\B\cdot\underline\B\cdot\vv - 
\frac{|\underline\B\cdot\vv|^2}{1-|\vv|^2}\vv\right).\end{equation*}

Introducing the customary axial vector $\B$ corresponding to $\underline\B$ we have $|\B|=|\underline\B|$, \
$\underline\B\cdot\vv=\vv\times\B$, \  $\underline\B\cdot\underline\B\cdot\vv=(\vv\times\B)\times\B$.
If the relative velocity is orthogonal to $\B$ then $\underline\B\cdot\underline\B\cdot\vv=-|\B|^2\vv$, 
$|\underline\B\cdot\vv|^2=|\B|^2|\vv|^2$
and the self-force relative to the inertial frame  
\beq-\frac1{4\pi}\frac{2e^4}{3m^2}\frac{|\B|^2\vv}{1-|\vv|^2}\enq 
brakes the motion; the same opposite force must be applied additionally to avoid braking and to ensure that the particle
moves on the prescribed orbit.

\subsection{On the LAD equation}

Besides the unjustified application of Gauss--Stokes theorem, of Taylor expansion and a doubtful limit to zero there is  
another, a fundamental error in usual derivations of the LAD equation which is clearly seen now.
\medskip

{\it The self-force $\frac{1}{4\pi}\frac{2e^2}{3}(\dot r\land\dddot r)\cdot\dot r$ is valid for a {\bf given} world line 
function $r$; therefore it is unjustified to put it in a Newtonian-like equation to obtain 
the LAD equation which would serve {\bf to determine} $r$.
Thus, the LAD equation is a {\bf misconception} and its pathological properties are not surprising}.
\medskip

The well known fundamental problem (\cite{Bargman}) is that both electrodynamics and mechanics in 
their known forms are {\bf theories of action}: 

-- \ the Maxwell equations define the electromagnetic field $\Ff$ produced by a {\bf given} 
world line function $r$ of a particle, 

-- \ the Newtonian equation defines the world line function $r$ of a particle in a {\bf given} force (e.g. 
the Lorentz force in an extraneous electromagnetic field $\F$),

\noindent and at present we have not a well working {\bf theory of interaction} which {\bf would define} both the 
electromagnetic field $\Ff$ and the world line function $r$ together.

To see the problem more closely, let us consider a point charge under the action of a given force  $\f$. Then according to the
general conception (see e.g. \cite{Gral-Har-Wald})  -- formulated in our language -- the Newtonian equation 
is replaced with the balance equation 
\beq\label{balance}-\Dd\cdot(\Tf_m[r] + \Tf_e[\Ff])+\f(r,\dot r)\la_{\ran r}=0\enq
where $\Tf_m[r]$ is the tensor Distribution of mechanical energy-momentum constructed 
somehow of $r$ and $\Tf_e[\Ff]$ is the tensor Distribution of electromagnetic energy-momentum constructed somehow of $\Ff$. 

It seems evident that $\Tf_m[r]=m\dot r\otm\dot r\la_{\ran r}$ and then the balance equation 
\beq\label{bala} m\ddot r\la_{\ran r} = \f(r,\dot r)\la_{\ran r} - \Dd\cdot\Tf_e[\Ff]\enq
together with the Maxwell equations 
\beq\label{maxw} \Dd\cdot\Ff=\dot r\la_{\ran r}, \qquad \Dd\land\Ff=0\enq
would describe that $r$ and $\Ff$ determine each other mutually.
  
It is not sure that there is a conveniently defined $\Tf_e[\Ff]$ with which
\eqref{bala} and \eqref{maxw} form a consistent system of equations.

It is sure, however, that $\Tf_e[\Ff]$ cannot be replaced with $\tm\T[r]$ which is valid for a {\bf given} $r$; the  
LAD `equation' is obtained by committing this replacement.

\subsection{Equation, equality}

At last, let the LAD `equation' be looked at from another point of view. Namely, 
we have to distinguish clearly between an equation and an equality, both denoted by the same symbol $=$.

An equation is a {\bf definition}: it defines a set (the set of its solutions).  

An equality is a {\bf statement}: it states that two sets are the same.

Let $S(e,m,\f)$ be the set of processes of a point charge $e$ with mass $m$ under the action of a force $\f$.

Because of the non-physical run-away solutions, the LAD `equation' does not define $S(e,m,\f)$, thus it is not an equation 
from a physical point of view.

The various attempts to find satisfactory conditions to exclude non-physical solutions suggest in this context that,
in fact, the LAD `equation' would be an equality, stating that $S(e,m,\f)$ is a subset of all the solutions.

Unfortunately, this is not true, either, as it is shown in the excellent paper \cite{eliezer}:
$\f$ being the Coulomb force of a point charge, no physical motion in radial direction is 
a solution of the LAD `equation'.

\subsection{An illustrative example}

\begin{figure}[hbt]
\centerline{\resizebox{0.65\textwidth}{!}{\includegraphics
{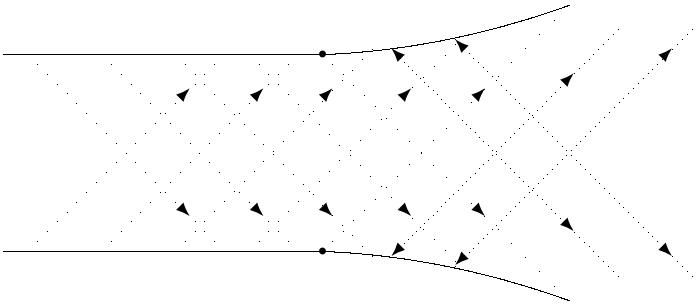}}}
\caption{\ Interaction of two point charges}
\label{fig:nemkolcs}
\end{figure}

The following example can illustrate the problem of describing the interaction. Let two particles with the same 
charge and mass rest in the space of a standard inertial frame. They repulse each other by the Coulomb forces which are compensated 
by some constraint to attain the rest. If the constraint is left off at an instant then the particles start to move. Figure 
\ref{fig:nemkolcs} shows what happens (time passes from the left to the right).

While resting, the particles  send Coulomb-actions to each other in the positive light cone, represented by 
sparsely dotted lines. At the instant when the constraint is left off, the acceleration of each particles
is determined by the Coulomb force of the other. Then for some time the acceleration of a particle is determined by 
the Coulomb force of the other particle and the radiation self-force, and then it sends some action, different
from the Coulomb force to the other particle; this is represented by the densely dotted lines. We have no 
formula for this acceleration and action. Later the situation becomes more complicated: it is not known 
which action and self-force give rise to an acceleration.

\subsection{Final remarks}

It is emphasized that the retarded field is meaningful only for a {\bf given} world line, therefore
it cannot be used for {\bf determining} the world line.

The various attempts to obtain a physically acceptable equation with continuum charge distributions instead of a point charge
suffer from the same fault of using the retarded field (and sometimes the advanced one, too).

\part*{Appendix}                                                        

\section{Measures}

Elementary notions of (vector) measures and integration by them can be found in \cite{Dinculeanu}.

The Lebesgue measure of an affine space is the measure given by the usual integration (customary for $\rr^n$).

The Lebesgue measure $\lambda_H$ of a submanifold $H$ in spacetime is given by the formula
$$\intl_H f \ d\lambda_H:= \intl_{\mathrm{Dom} p} f(p(\xi))\sqrt{|\det(\Dp p[\xi]^*\Dp p[\xi])|}\ d\xi$$
where $p$ is an arbitrary parameterization of the submanifold i.e. a continuously differentiable injective map 
defined in an affine space, having $H$ as its range and $\Dp p[\xi]$, the derivative of $p$ at $\xi$ is injective for all $\xi$.

In particular, for a world line function $r$, the Lebesgue measure of the corresponding world line is $\lambda_{\ran r}$ 
which gives the integration along the world line. In particular, it is considered a Distribution in spacetime, 
\beq\label{ranr}\left(\lambda_{\ran r}\mid\phi\right)=\intl_\I \phi(r(\s))\ d\s\enq
for arbitrary  test functions $\phi$.

\section{Three dimensional integration}\label{4pi3}

Let $\uu_0$ be an absolute velocity in spacetime and $\Sd_0$ the three dimensional Euclidean vector space Lorentz orthogonal 
to $\uu_0$; $S_1(\0)$ is the unit sphere around zero, its elements are denoted by $\nb_0$.

In the usual parameterization  given by the angles $\vartheta,\vf$ relative to an orthonormal bases, 
the Lebesgue measure of $S_1(\0)$ is obtained by the symbolic formula $d\nb_0\sim \sin\vartheta\ d\vartheta\ d\vf$ 
with which we easily find the equalities 

\beq\label{n1} \intl_{S_1(\0)} \nb_0 \ d\nb_0=0, \qquad 
\intl_{S_1(\0)} \nb_0\otimes\nb_0\otimes\nb_0 \ d\nb_0=0,\enq
\beq\label{nn} \intl_{S_1(\0)} \nb_0\otimes\nb_0 \ d\nb_0=\frac{4\pi}{3}(\1 + \uu_0\otm\uu_0)\enq
where $\1$ is the identity map of spacetime vectors; and so on, the integral of an even tensor products is a non-zero tensor,
the integral of an odd tensor products is zero.

The radial coordinates of $\q\in\Sd_0$ are 
$$ \rho(\q):=|\q|\ge0, \quad \nb_0(\q):=\frac{\q}{|\q|}\in S_1(\0).$$

Then the radial parameterization of $\Sd_0$ is 

$$ \label{radpar}(\rho,\nb_0)\mapsto \rho\nb_0.$$

We applied the usual ambiguity that $\rho$ and $\nb_0$ denote functions in some respect and variables 
in an other respect.

The other well known integral formula
\beq\label{radint1}\intl_{\Sd_0} f(\q)\ d\q =\intl_0^\infty \rho^2\intl_{S_1(\0)}f(\rho\nb_0)\ d\nb_0\ d\rho\enq
will be often used, too.

\section{Pole taming}\label{polet}

There is a well defined method to attach a Distribution to a locally non-integrable function, having a pole
singularity in a Euclidean space. Originally the method was called regularization (\cite{Gelf-Shil}).
Since the Distribution obtained in this way is not regular, I call the method pole taming\footnote{\'Aron Szab\'o 
proposed this name}.

Only a special case is necessary for us. 

For the function $\rho(\q):=|\q|$ \ $(\q\in\Sd_0)$, 
$\frac1{\rho^{2+m}}$ is not locally integrable if $m>0$. For $m$ being an even positive integer, a Distribution 
can be attached to it by pole taming defined as follows.

For a test function $\psi$ let $\q\mapsto T_\psi^{(m-1)}(\q)$ denote the Taylor polynomial of order $m-1$ at zero.
In a neighbourhood of zero, $\psi-T_\psi^{(m-1)}$ is a continuous function of order $\rho^m$, therefore  
$\frac{\psi-T_\psi^{(m-1)}}{\rho^{2+m}}$ is integrable there. Outside the support of $\psi$, however, the term 
containing the $(m-1)$-th derivative is not integrable.
 
$m$ is even; in the radial parameterization \eqref{radpar}, $\Dp^{(m-1)}\psi(\0)(\rho\nb_0,\dots,\rho\nb_0)$ is a 
linear function of an odd tensor product of $\nb_0$, thus,
integrating in the order given by \eqref{radint1}, this term will drop out and an integrable function remains.

We repeat for avoiding misundertanding: it is known that, for an integrable function, the order of integration by 
$\nb_0$ and $\rho$ can be interchanged in \eqref{radint1}. If a function is not integrable, it can occur that in one of the 
orders the integral exists (but in the other order not). Here we take the advantage that the integral exists in the 
given order.
In this way the pole taming of $\frac1{\rho^{2+m}}$ results in the Distribution defined by 
\beq\label{tomp} \left(\tm\frac1{\rho^{2+m}}\Bigm|\psi\right):=
\intl_0^\infty\left(\rho^2\intl_{S_1(\0)}\frac{\psi(\rho\nb_0) - T_\psi^{(m-1)}(\rho\nb_0)}{\rho^{2+m}}\ d\nb_0\right)\ d\rho\ !\enq
where the exclamation mark calls attention to that the order of integration cannot be interchanged.

For $m=0$ we mean the pole taming formally by taking the corresponding regular Distribution.

Later, for the sake of brevity, we also use the formula
\beq\label{tompi} \left(\tm\frac1{\rho^{2+m}}\Bigm|\psi\right):=
\intl_{\S_0}\frac{\psi(\q) - T_\psi^{(m-1)}(\q)}{|\q|^{2+m}}\ d\q\ !!\enq
where the double exclamation mark calls attention to that the integral must be taken in radial parameterization and in the 
given order.

We have to extend the notion of pole taming as follows. For a $k$-th tensor power $\nb_0^{\otm k}:=\nb_0\otm\nb_0\otm\dots\nb_0$ 
we put
\beq\label{npol}\left(\tm\frac{\nb_0^{\otm k}}{\rho^{2+m}}\Bigm|\psi\right):=
\intl_0^\infty \rho^2\left(\intl_{S_1(\0)}\frac{\nb_0^{\otm k}
\bigl(\psi(\rho\nb_0) - T_\psi^{(m-1)}(\rho\nb_0\bigr)}{\rho^{2+m}}\ d\nb_0\right)\ d\rho\ !\enq
which makes sense if $m-1+k$ is odd; therefore 

-- \ if $m$ is even then $k$ must be even, too,

-- \ if $m$ is odd then $k$ must be odd, too.

\section{World rolls and world tubes \\ around a world line}\label{roll}

Let $r$ be a twice continuously differentiable world line function. The Lorentz boost (see \cite{Matolcsi})
$$ \Ll(\s):=\1 +\frac{(\dot r(0) + \dot r(\s))\otimes(\dot r(0) + \dot r(\s))}{1 -\dot r(0)\cdot\dot r(\s)} -
2\dot r(\s)\otimes\dot r(0)$$
maps $\dot r(0)$ into $\dot r(\s)$. 

Further, we use the notations $\uu_0:=\dot r(0)$ and $\Sd_0$ for the linear subspace consisting of 
vectors Lorentz orthogonal to $\uu_0$; then $\{\Ll(s)\cdot\q\mid \q\in\Sd_0\}$ is the subspace consisting of
vectors Lorentz orthogonal to $\dot r(\s)$. Then

\beq\label{param} p:\I\times\Sd_0\to \aM, \qquad (\s,\q)\mapsto r(\s) + \Ll(\s)\cdot\q\enq                        
is a parameterization of a neighbourhood of the world line.

This is a reformulation of a usual setting: $(\s,\q)$ correspond to the so called Fermi coordinates.

To be precise, the following details are necessary. 
$p$ is evidently continuously differentiable  and its derivative 
\beq\label{Pder} \Dp p[\s,\q]= \Bigl(\dot r(\s) + \dot\Ll(\s)\cdot\q \ \ \Ll(\s)|_{\Sd_0}\Bigr)\enq
($\Ll(\s)|_{\Sd_0}$ is the restriction of $\Ll(\s)$ to $\Sd_0$) is injective for all $\s$ and for $\q=\0$, thus, 
according to the inverse function theorem and the fact that 
$p(\s,\0)=r(\s)$, there is an open subset in $\I\times\Sd_0$ which contains $\I\times\{\0\}$ where 
both $p$ and all the values $\Dp p[\ \cdot \ ] $ are injective. 

Let $I$ be a bounded and closed time interval. Then $\{r(\s)\mid \s\in I\}$ is a compact set and each of its points 
has a neighbourhood in the range of the parameterization; thus, $\{r(\s)\mid \s\in I\}$ can be covered 
by finite many such neighbourhoods. As a consequence,   
there is an $R_I>0$ in such a way that the part $\{r(\s)\mid \s\in I\}$ of the world line
is contained in 
$$ H_{I,R}:= \{r(\s) + \Ll(\s)\cdot\q\mid \s\in I, |\q|<R\}$$
for all $0<R\le R_I$. 

$H_{I,R}$ is called a {\bf world roll} of length $I$ and radius $R$. 
With the notations
\beq\label{rno}\rho(\q):=|\Ll(\s)\cdot\q|=|\q|,  \qquad \nb(\s,\q):=\frac{\Ll(\s)\cdot\q}{\rho(\q)}, 
\qquad \nb_0(\q):=\frac{\q}{\rho(\q)},\enq
the parameterization of $H_{I,R}$ can be rewritten in the form
\beq\label{rron} r(\s) + \rho(\q)\nb(\s,\q) = r(\s) + \Ll(s)\cdot\nb_0(\q).\enq

The cylinder around $H_{I,R}$,
$$ P_{I,R}:=\{r(\s) + \Ll(\s)\cdot\q \mid \s\in I,\ |\q|=R\}$$
is called the corresponding {\bf world tube}.

The tangent space of $P_{I,R}$ at the point $r(\s) +\Ll(\s)\cdot\q$ is 
$$ \{\bigl(\dot r(\s) + \dot\Ll(\s)\cdot\q\bigr)\ts + \Ll(\s)\cdot\hh\mid \ts\in\I, \ \hh\in\Sd_0, \ \hh\cdot\q=0\}.$$

For $\q\in\Sd_0$ we have 
$(\Ll(\s)\cdot\q)\cdot \dot r(\s)=\0$, $(\Ll(\s)\cdot\q)\cdot(\Ll(\s)\cdot\hh)=\q\cdot\hh=0$ and
$(\dot\Ll(\s)\cdot\q)\cdot(\Ll(\s)\cdot\q) = \frac1{2}\frac{\pt}{\pt\s}|\Ll(\s)\cdot\q|^2=0$; thus,
$$\Bigl(\bigl(\dot r(\s) + \dot\Ll(\s)\cdot\q\bigr)\ts + \Ll(\s)\cdot\hh\Bigr)\cdot (\Ll(\s)\cdot\q)=0$$ 
and we find that
the outward normal vector of $P_{I,R}$ at the point $r(\s) + \rho(\q)\nb(\s,\q)$ is
\beq\label{palnorm}\nb(\s,\q).\enq

\section{Integration in world rolls}

Integrals in a world roll will be computed in the parameterization \eqref{param} for which we have

\beq\label{rna} \sqrt{|\det\Dp p[\s,\q]^*\Dp p[\s,\q]|}=1+\rho(\q)\nb(\s,\q)\cdot\ddot r(\s).\enq

To show it, for the sake of brevity, we introduce the notation 
$$\z(\s,\q):=\dot r(\s) + \dot\Ll(\s)\cdot\q.$$

Then from \eqref{Pder} we get                                                            
\begin{align} \Dp p[\s,\q]^*\Dp p[\s,\q] &= \begin{pmatrix} \z(\s,\q) \\ \bigl(\Ll(\s)|_{\Sd_0}\bigr)^*\end{pmatrix}
\begin{matrix}\cdot\begin{pmatrix} \z(\s,\q) & \Ll(\s)|_{\Sd_0}\end{pmatrix} \\ &  \end{matrix}= \\
&=\label{blokk} \begin{pmatrix} \z(\s,\q)\cdot\z(\s,\q) & \z(\s,\q)\cdot\Ll|(\s)_{\Sd_0} \\ 
(\Ll(\s)|_{\Sd_0})^*\cdot\z(\s,\q) & \1_{\Sd_0}\end{pmatrix}.\end{align}

Omitting the variables for the sake of perspicuity, we can write $\Ll|_{\Sd_0}=\Ll\cdot(\1+\uu_0\otimes\uu_0)$, thus
$(\Ll|_{\Sd_0})^*=(\1+\uu_0\otimes\uu_0)\cdot\Ll^*$; accordingly,
$(\1+\uu_0\otimes\uu_0)\cdot\Ll^*\cdot\z=\Ll^*\cdot\z + \uu_0((\uu_0\cdot\Ll^*)\cdot\z$ which equals
$\z\cdot\Ll + \uu_0(\dot r\cdot\z)=\z\cdot(\Ll\cdot(\1+\uu_0\otimes\uu_0))$.

The block matrix \eqref{blokk} is symmetric, its determinant is $\z\cdot\z - (\z\cdot\Ll|_{\Sd_0})\cdot((\Ll|_{\Sd_0})^*\cdot\z)$.
The second term here equals $(\z\cdot\Ll)\cdot(\z\cdot\Ll) + 
2(\z\cdot\Ll)\cdot\uu_0(\dot r\cdot\uu_0)-(\dot r\cdot\uu_0)^2=\z\cdot\z + (\dot r\cdot\uu_0)^2,$
thus we have
$$ \det(\Dp p^*\Dp p)=-(\dot r\cdot \z)^2 = -(1 + \rho\nb\cdot\ddot r)^2;$$
the last equality comes from
$$ \dot r(\s)\cdot\z(\s,\q) =-1 +\dot r(\s)\cdot \dot\Ll(\s)\cdot\q=-1 + \frac{\pt}{\pt\s}\bigl(\dot r(\s)\cdot\Ll(s)\cdot\q\bigr) -
\ddot r(\s)\cdot\Ll(\s)\cdot\q $$
and the quantity in the parenthesis in the middle term on the right hand side is zero.
                                                                                          
Finally, if $\rho(\q)$ is `sufficiently small' then $1 + \rho(\q)\nb(\s,\q)\cdot\ddot r(\s)$ is positive, thus
it is positive for all the possible $(\s,\q)$ because $\Dp p[\s,\q]$ is injective, the determinant cannot be zero. 

Further, because of $\nb(\s,\q)=\Ll(\s)\nb_0(\q)$, using the radial parameterization \eqref{radpar} of $\Sd_0$ we have
\beq\label{ezazz} \intl_{H_{I,R}} \dots (x)\ dx =\intl_0^R \rho^2\intl_{S_1(0)} \dots 
\bigl(r(\s) + \rho\Ll(\s)\nb_0\bigr)(1 + \rho\ddot r(\s)\cdot\Ll(\s))\ d\nb_0 \ d\rho.\enq

\section{Acknowledgement} I am grateful to T.Gruber for checking all the not too simple formulas and to P.V\'an for helping me
to compose this article.


\begin{thebibliography}{99}

\bibitem{Matolcsi} T. Matolcsi: {\it Spacetime without Reference Frames}, Minkowski Institute Press, 2020

\bibitem{Jackson} J. D. Jackson: {\it Classical Electrodynamics} 3rd Ed. Wiley, New York, 1998

\bibitem{Groot} S. R. de Groot -- L. G. Suttorp: {\it Foundations of Electrodynamics}, 
North-Holland, 1972

\bibitem{Spohn} H.Spohn. J.Phys.A {\bf 44} (2011)

\bibitem{Rohrlich1} F.Rohrlich Am.J.Phys. {\bf 65} (1997) p1051


\bibitem{Rohrlich2} F. Rohrlich: {\it Classical Charged Particles} 3rd Ed. World Scientific, 2007

\bibitem{Bay-Hus} W.E.Baylis--J.Hushilt Phys.Rev. D {\bf 13} (1976)

\bibitem{Bild-Deck-Ruhl} C.Bild--D.A. Deckert--H.Ruhl. Phys. Rev. D, {\bf 99} (2019)

\bibitem{Gral-Har-Wald} S.E.Gralla--A,I.Harte--R.M.Wald Phys. Rev. D {\bf 80} (2009)

\bibitem{Taylor} J.G.Taylor: {\it  Classical electrodynamics as a distribution theory I-II}, 
Mathematical Proceedings of the Cambridge Philosophical Society, 
52/1, (1956), p119--134 and 54/2, (1958), p258--264.

\bibitem{Spohnc} H. Spohn, Europhys, Lett. {\bf 50/3} (2000), p287

\bibitem{lordir} T. Matolcsi -- T. Fülöp -- M. Weiner, Modern Physics Letter A {\bf 32} (2017), p1750147

\bibitem{Rowe} E.G.P Rowe Phys. Rev. D {\bf 18} (1978), p3639

\bibitem{MoPap} T.C.Mo -- C.H.Papas, Phys. Rev. D {\bf 4} (1971), 3566

\bibitem{steane} A.M.Steane, Am. J. Phys {\bf 83} (2015) p.703


\bibitem{Gelf-Shil} I. M. Gelfand -- G. E. Shilov: {\it Generalized Functions I}, Acad. Press, 1964

\bibitem{Demidov} A. D. Demidov: {\it Generalized Functions in Mathematical Physics}, 
Nova Sciences, 2001

\bibitem{Dijk} G. van Dijk: {\it Distribution Theory},  De Gruyter Graduate Lectures, 2010

\bibitem{Horvath} J. Horv\'ath: {\it Topological Vector Spacaes and Distributions}, 
Dover Publication, 2012

\bibitem{matvan} T.Matolcsi -- P.V\'an (2020) arXiv 2010.04940

\bibitem{eliezer} C.J.Eliezer Math. Proc. Cambridge Phil. Soc. {\bf 39}. Nr 3. (1943), p173 

\bibitem{Boulware} D.G.Boulware, Ann.Phys {\bf 124} (1980) p.169

\bibitem{Eriksen} F.Eriksen -- \/O.Gr\/on, Ann. Phys. {\bf 286} (2000) p.320

\bibitem{Bargman} P.G.Bargman {\it Basic Theories of Physics}, New York:Dover 1962

\bibitem{Dinculeanu} N. Dinculeanu: {\it Vector Measures}, Pergamon Press, 1967

\bibitem{Born} M. Born - E. Wolf: {\it Principles of Optics}, Pergamon Press, 1970
\end{thebibliography}
\end{document}